\def \SAIT #1 #2 {{\em Mem.\ Soc.\ Astron.\ It.\/} {\bf #1}, #2}
\def \MESS #1 #2 {{\em The Messenger\/} {\bf #1}, #2}
\def \ASTRNACH #1 #2 {{\em Astron. Nach.\/} {\bf #1}, #2}
\def \AAP #1 #2 {{\em Astron. Astrophys.\/} {\bf #1}, #2}
\def \AAL #1 #2 {{\em Astron. Astrophys. Lett.\/} {\bf #1}, L#2}
\def \AAR #1 #2 {{\em Astron. Astrophys. Rev.\/} {\bf #1}, #2}
\def \AAS #1 #2 {{\em Astron. Astrophys. Suppl. Ser.\/} {\bf #1}, #2}
\def \AJ #1 #2 {{\em Astron. J.\/} {\bf #1}, #2}
\def \ANNREV #1 #2 {{\em Ann. Rev. Astron. Astrophys.\/} {\bf #1}, #2}
\def \APJ #1 #2 {{\em Astrophys. J.\/} {\bf #1}, #2}
\def \APJL #1 #2 {{\em Astrophys. J. Lett.\/} {\bf #1}, L#2}
\def \APJS #1 #2 {{\em Astrophys. J. Suppl.\/} {\bf #1}, #2}
\def \APSS #1 #2 {{\em Astrophys. Space Sci.\/} {\bf #1}, #2}
\def \ASR #1 #2 {{\em Adv. Space Res.\/} {\bf #1}, #2}
\def \BAIC #1 #2 {{\em Bull. Astron. Inst. Czechosl.\/} {\bf #1}, #2}
\def \JSQRT #1 #2 {{\em J. Quant. Spectrosc. Radiat. Transfer\/} {\bf #1}, #2}
\def \MN #1 #2 {{\em Mon. Not. R. Astr. Soc.\/} {\bf #1}, #2}
\def \MEM #1 #2 {{\em Mem. R. Astr. Soc.\/} {\bf #1}, #2}
\def \PLR #1 #2 {{\em Phys. Lett. Rev.\/} {\bf #1}, #2}
\def \PASJ #1 #2 {{\em Publ. Astron. Soc. Japan\/} {\bf #1}, #2}
\def \PASP #1 #2 {{\em Publ. Astr. Soc. Pacific\/} {\bf #1}, #2}
\def \NAT #1 #2 {{\em Nature\/} {\bf #1}, #2}

\documentstyle{memsait}
\input epsf.sty
\begin{opening}
\title{MULTIFREQUENCY STUDIES OF GAMMA-RAY BURSTS: 
TOWARDS THE UNDERSTANDING OF THE MYSTERY}
\author{ALBERTO J. CASTRO-TIRADO$^{1,2}$}
\institute{$^1$Laboratorio de Astrof\'{\i}sica Espacial
           y F\'{\i}sica Fundamental (LAEFF-INTA),
           Apdo. 50727, E-28080, Madrid, Spain \\
$^2$Instituto de Astrof\'{\i}sica de Andaluc\'{\i}a (IAA-CSIC),
           Apdo. 03004, E-18080, Granada, Spain}
\date{} 
\end{opening}

\begin{document}

\oddpagefooter{}{}{} 
\evenpagefooter{}{}{} 
\ 
\bigskip

\begin{abstract}
GRBs have remained a puzzle
for many high--energy astrophysicists since their discovery in 1967.
With the advent of the X--ray satellites {\it BeppoSAX} and {\it RossiXTE}, 
it has been possible to carry out deep multi-wavelength observations of 
the counterparts associated with the GRBs just within a few hours of 
occurence, thanks to the observation of the fading X-ray emission that 
follows the more energetic gamma-ray photons once the GRB event has ended. 
The fact that this emission (the afterglow) extends at longer wavelengths, 
has led to the discovery of the first optical/IR/radio counterparts in 
1997-99, greatly improving our understanding of these sources.
Now it is widely accepted that GRBs originate at cosmological
distances. The observed afterglow satisfies the predictions of the 
"standard" relativistic fireball model, and the central engines that power 
these extraordinary events are thought to be the collapse of massive stars
or the merging of compact objects. 
\end{abstract}

\section{Introduction.}

In 1967-73, the four VELA spacecraft (named after the spanish verb 
{\it velar}, to keep watch), that where originally designed for verifying 
whether the former Soviet Union abided by the Limited Nuclear Test Ban 
Treaty of 1963, observed 16 peculiarly strong events 
(Klebesadel, Strong and Olson 1973, Bonnell and Klebesadel 1996).
On the basis of 
arrival time differences, it was determined that they were related neither
to the Earth nor to the Sun, but they were of cosmic origin. Therefore they
were named cosmic Gamma-Ray Bursts (GRBs hereafter).\par

GRBs appear as brief flashes of cosmic high energy photons, emitting the 
bulk of their energy above $\approx$ 0.1 MeV. 
The KONUS instrument on {\it Venera 11} and {\it 12} gave the first 
indication that GRB sources were isotropically distributed in the sky 
(Mazets et al. 1981, Atteia et al. 1987).  Based on a much larger sample, this 
result was nicely confirmed by BATSE on board the {\it CGRO} satellite 
(Meegan et al. 1992). 
In general, there was no evidence of periodicity in the 
time histories of GRBs.  However there was indication of a bimodal 
distribution of burst durations, with
$\sim$25\% of bursts having durations around 0.2 s and $\sim$75\% with 
durations around 30 s. 
A deficiency of weak events was noticed in the log $N$-log $S$ diagram,
as the GRB distribution deviates from the -3/2 slope of the straight line 
expected
for an homogeneous distribution of sources assuming an Euclidean geometry. 
However, the GRB distance scale had to 
remain unknown for 30 years.  A comprenhensive review of these observational 
characteristics can be seen in Fishman and Meegan (1995).

\section{The search and detection of counterparts at other wavelengths}

It was well known that an important clue for solving the GRB puzzle was going
to be the detection of transient emission -at longer wavelengths- associated 
with the bursts.
A review on the unsuccessful search for counterparts prior to 1997 
can be seen in Castro-Tirado (1998) and references therein.
Here I will present some results concerning six selected bursts detected by 
the {\it BeppoSAX} ({\it BSAX}) and {\it RossiXTE} ({\it RXTE}) satellites in 
1996-99 and their impact on the current understanding on the physics of GRBs.

\subsection{GRB 970228}
Thanks to {\it BSAX}, it was possible on 28 Feb 
1997 to detect the first {\it clear} evidence of a long X-ray tail 
 -the X-ray afterglow- following GRB 970228. A previously unknown 
X-ray source was seen to vary by a factor of 20 on a 3 days timescale. 
The X-ray fluence was $\sim$ 40 \% of the gamma-ray fluence, as reported 
by Costa et al. (1997), implying that the X-ray 
afterglow was not only the low-energy tail of the GRB, but also a significant 
channel of energy dissipation of the event on a completely different timescale.
Another important result was the non-thermal origin of the burst radiation 
and of the X-ray afterglow (Frontera et al. 1998). The precise X-ray 
position (1$^{\prime}$) led to the discovery of the first optical transient 
(or optical afterglow, OA) associated to a GRB, identified on 28 Feb 1997, 
20 hr after the event (Groot et al. 1997, van Paradijs et al. 1997).
The OA was afterwards found on earlier images taken by Pedichini et al. 
(1998) and Guarnieri et al. (1997), 
in the rising phase of the light curve. The maximum was reached $\sim$ 20 hr 
after the event (V $\sim$ 21.3), and followed by a power-law decay  
F $\propto$ t$^{-1.2}$ (Galama et al. 1997, Bartolini et al. 1998). 
An extended source was seen at the OA position since the very beginning 
by ground-based and {\it HST} observations (van Paradijs et 1997, 
Sahu et al. 1997).
New {\it HST} observations taken 6 months after the event were reported by 
Fruchter et al. (1997) and both the OA (at V = 28) and the extended source 
(V = 25.6) were seen. 
The extended object surrounding the point-source was  interpreted as a 
galaxy, according to the similarities (apparent size, magnitude) with 
objects in the {\it HST} Deep Field. Finally, after two years of work, 
the redshift of this object has been determined as z = 0.695 (Djorgovski et
al. 1999a), confiming its extragalactic nature and implying a star-forming 
rate comparable to other galaxies at similar redshifts. Reichart (1999) 
proposed a type Ib/c supernova lies "behind" the GRB, overtaking the light 
curve two weeks after. This fact seems to be confirmed by the work of Galama 
et al. (2000).  

\begin{figure}[t]
\epsfysize=6cm 
\hspace{3.5cm}\epsfbox{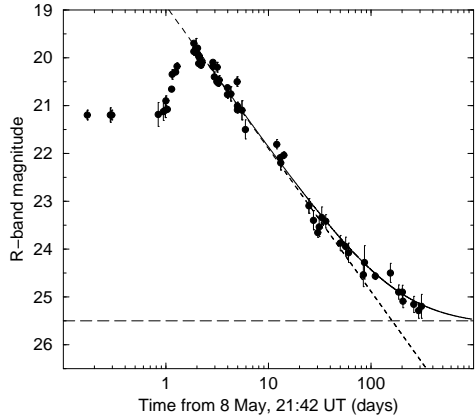} 
\caption[h]{The R-band light curve of the GRB 970508 optical afterglow, from
           data quoted in this paper. The dotted like is the contribution of 
           the GRB afterglow itself, following F $\propto$ t$^{-1.19}$ two 
           days after the burst. The horizontal dashed line is the R = 25.5 
           host galaxy, whereas the solid line is the contribution of both 
           (afterglow plus host galaxy). From Castro-Tirado and Gorosabel 
           (1999).}
\end{figure}

\subsection{GRB 970508}
The second OA associated to a GRB was discovered by Bond (1997) 
within the GRB 970508 error box, and observed 3 hr after the burst 
in unfiltered images (Pedersen et al. 1998). The optical light 
curve reached a peak in two days (R = 19.7, Castro-Tirado et al. 1998a, 
Djorgovski et al. 1997, Galama et al. 1998a) and 
was followed by a power-law decay F $\propto$ t$^{-1.2}$. 
Optical spectroscopy obtained during the maximum allowed a 
direct determination of a lower limit for the redshift of GRB 970805 
($z \geq 0.835$), implying E $\geq$ 7 $\times$ 10$^{51}$ erg 
and was the first proof that GRB
sources lie at cosmological distances (Metzger et al. 1997). The 
flattening of the decay in late August 1997 (Pedersen et al. 1998, Sokolov et
al. 1998) revealed the contribution of a constant brightness source 
-the host galaxy- that was revealed in late-time imaging obtained
in 1998 (Bloom et al. 1998, Castro-Tirado et al. 1998b, 
Zharikov et al. 1998). See Fig. 1. 
The maximum observed 1-day after the event has not been detected in
other GRBs and it was interpreted by a delayed energy injection or
by an axially symmetric jet surrounded by a less energetic outflow
(Panaitescu et al. 1998).
The luminosity of the galaxy is well below the knee of the galaxy luminosity 
function, L $\approx$ 0.12 $L^{*}$, and the detection of deep 
Mg I absorption (during the bursting episode) and strong [O II] 3727 $\rm \AA$ 
emission (the latter mainly arising in H II regions within the host galaxy) 
confirmed $z$ = 0.835 and suggested that the host could be a normal dwarf 
galaxy (Pian et al. 1998), with a star formation 
rate (SFR) of $\sim$ 1.0 $M_{\odot}$ year$^{-1}$ (Bloom et al. 1998).
Prompt VLA observations of the GRB 
970508 error box allowed detection of a variable radio source at 1.4, 4.8 
and 8.4 GHz, the first radiocounterpart ever found for a GRB (Frail et al. 
1997). The fluctuations could be the result of 
strong scattering by the irregularities in the ionized Galactic 
interstellar gas, with the damping of the fluctuations with time indicating 
that the source expanded to a significantly larger size.
However VLBI observations did not resolve the object (Taylor et al. 1997).
The transient was also detected at
15 GHz (Pooley and Green 1997) and as a continuum point source at 86 GHz with 
the IRAM PdBI on 19-21 May 1997 (Bremer et al. 1998).
A Fe K$\alpha$ line redshifted at $z$ = 0.835 in the X-ray 
afterglow spectrum (Piro et al. 1999) was attributed to a thick torus 
surrouding the central engine (M\'esz\'aros and Rees 1998).
GRB 970508 is the best observed afterglow so far. The broad band
spectrum (see Fig. 2) is nicely explained by the standard relativistic
blast wave model (Wijers and Galama 1999).

\begin{figure}[t]
\epsfysize=6cm 
\hspace{3.5cm}\epsfbox{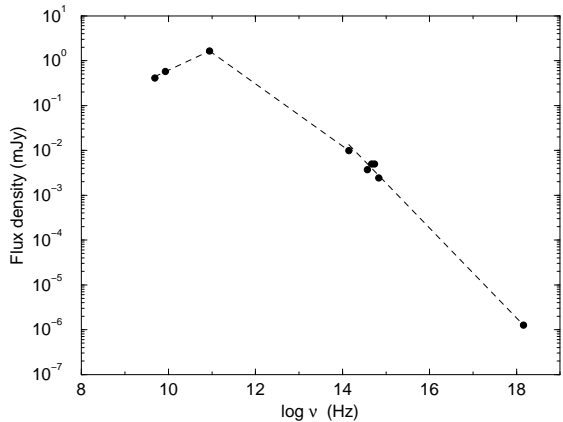} 
\caption[h]{The multiwavelength spectrum of GRB 970508, on May 22, 1997.
         Adapted from Gorosabel (1999). See also Wijers and Galama (1999).} 
\end{figure}

\subsection{GRB 970828}
This burst was detected by {\it RXTE} (Remillard et al. 1997) and was followed
up by {\it ASCA} and {\it ROSAT} (Murakami et al. 1997, Greiner et al. 1997).
 The fact that no optical counterpart down to R = 23.8 was detected between 
4 hr and 8 days after the event, could support the idea that the non-detection
was due to photoelectric absorption (Groot et al. 1998). The X-ray spectrum 
as seen by {\it ASCA} is strongly absorbed, suggesting that the event 
occurred in a dense medium. An excess at 6.7
keV was foud by {\it ASCA} in the X-ray afterglow spectrum. If this is due
to highly ionized Fe, then $z$ $\sim$ 0.33 (Yoshida et al. 1999) and the host
would be another dwarf galaxy (Gorosabel 1999). However, if the transient
radiosource detected with the VLA is indeed associated to the event, the
galaxy would be at a redshift of 0.96 (Frail et al. 2000).

\subsection{GRB 980425}
A peculiar type Ib/c supernova (SN 1998bw) was found in the WFC error box for 
this soft GRB (Galama et al. 1998b). The SN lies in the galaxy ESO 184-82 (at 
$z$ = 0.0085). The fact that the SN event occurred within $\pm$ 1 day 
of the GRB event, together with the relativistic expansion speed derived from 
the radio observation (Kulkarni et al. 1998a) strengths such a relationship. 
In that case, the total energy released would be 8 $\times$ 10$^{47}$ erg 
which is about $\sim$ 10$^{5}$ smaller than for "classical" GRBs.
The fact that a fading X-ray source -as in {\it all} the previous
cases- unrelated to the SN was detected by {\it BSAX} in the 
GRB error box (Pian et al. 1999, Piro et al. 1998) 
cast some doubts on the SN/GRB association (Graziani et al. 1999).
Although a deeper X-ray observation is pending, there is a general 
agreement now that the SN 1998bw/GRB 980425 relationship is real.

\begin{figure}[t]
\epsfysize=6cm 
\hspace{3.5cm}\epsfbox{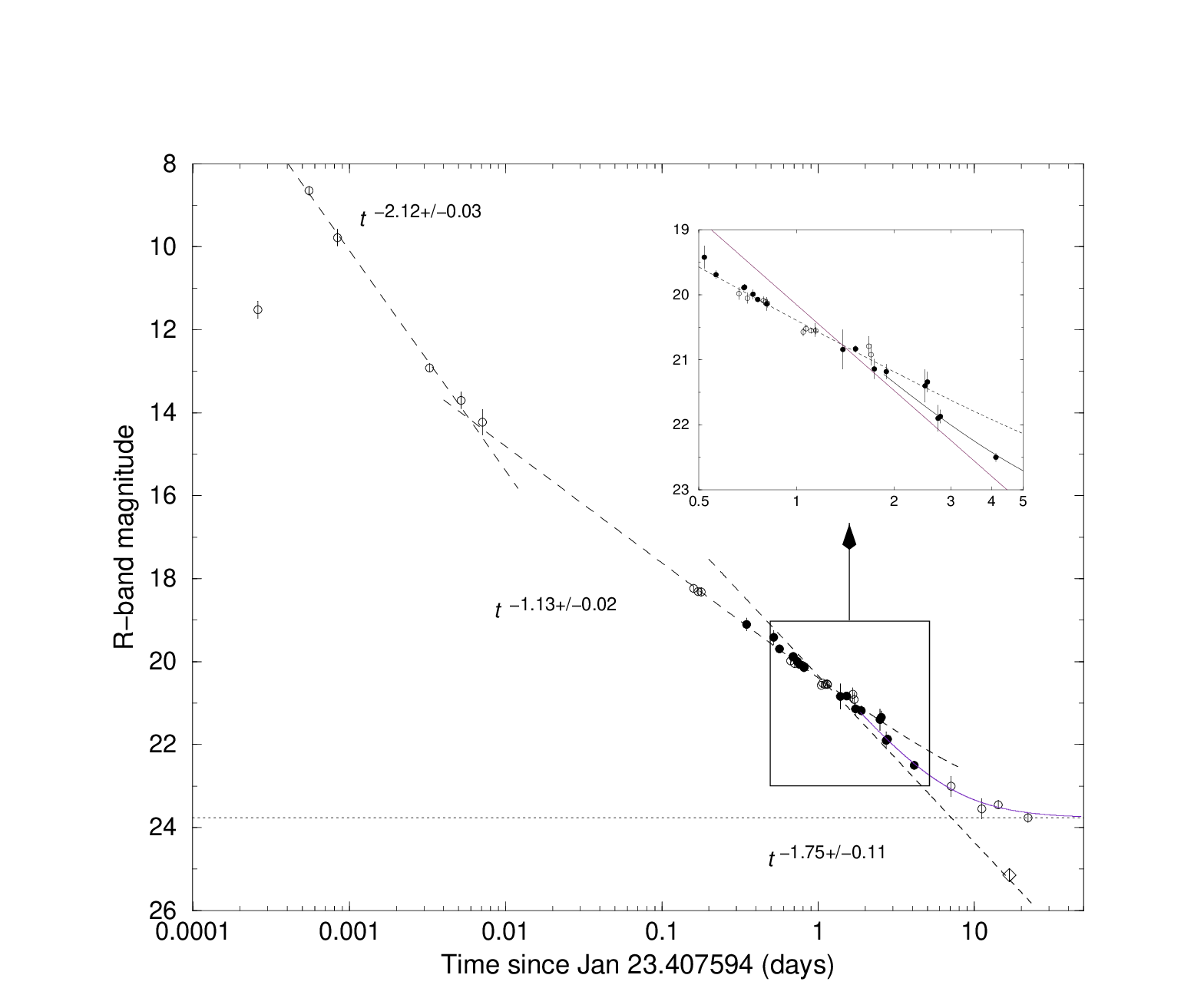} 
\caption[h]{ The R-band light-curve of the GRB~990123 optical transient.
 Based on our observations (filled circles) and other data reported elsewhere 
 (empty circles) ({\it 11,23}). 
 The doted line is the contribution of the  
 underlying galaxy, with R $\sim$ 23.77 $\pm$ 0.10, from ({\it 21}). 
 The three dashed lines are the contribution of the OA, following 
 F $\propto$ $t^{\delta}$ with $\delta$ = $-$2.12 up to $\sim$ 10 min, 
 $\delta$ = $-$1.13 up to $\sim$ 1.5 d, and
 $\delta$ = $-$1.75 after that time. The solid line, only drawn after 1.5 d 
 for clarity is the total observed flux (OA + galaxy).
 From Castro-Tirado et al. (1999).} 
\end{figure}

\subsection{GRB 990123}
This is the first for which contemporaneous optical emission was
found simultaneous to the gamma-ray burst, reaching V $\sim$ 9  (Akerloff
et al. 1999). This optical flash did not track the gamma-rays and did not
fit the extrapolation of theSAX and BATSE spectra towards longer wavelengths. 
This optical emission was interpreted as the signature of a reverse shock
moving into the ejecta (Sari and Piran 1999). 
A brief radiotransient was also detected (Frail et al. 1999a)
coincident with the optical counterpart (Odewahn et al. 1999) and  
spectrocopy indicated a redshift z = 1.599 (Kulkarni et al. 1999, Andersen 
et al. 1999). A break observed in the light curve $\sim$ 1.5
days after the high energy event suggested the presence of a beamed outflow
(Castro-Tirado et al. 1999, Fruchter et al. 1999, Kulkarni et al. 1999). 
See Fig. 3. A weak magnetic field 
in the forward shock region could account for the observed multiwavelength 
spectrum in contrast to the high-field for GRB 970508 and it seems that
the emission from the three regions was first seen in this event (Galama et 
al. 1999a): the internal, reverse and forward shocks. 

\subsection{GRB 990510}
Following the BSAX/WFC detection, an optical counterpart was reported by 
Vreeswijk et al. (1999a). The acromatic break seen in the light curve was
also interpreted as a jet, with a model yielding an opening angle of 
0.08 and a beaming factor of 300 (Harrison et al. 1999).
This is the first burst for which polarized 
optical emission was detected ($\Pi$ = 1.7 $\pm$ 0.2 \%), by means of an 
observation performed $\sim$18.5 hr after the event (Covino et al. 1999)
and later on (Wijers et al. 1999). 
This confirms the synchrotron origin of the blast wave itself and represents
the second case for a jet-like outflow (Stanek et al. 1999). 

Further X-ray afterglows were observed by {\it BSAX} and {\it RXTE} in 
1997-99. The optical afterglow of GRB 980326 was suggested to resemble a
SN at late times (Castro-Tirado and Gorosabel 1999) and indeed the late 
time light curve of GRB 980326 was explained by an underlying SN 1998bw SN 
at redshift of around unity (Bloom et al. 1999), thus strenghtening a possible
SN-GRB connection. 

Exponents for the power-law decay in the X-rays and in the optical are in the 
range $\alpha$ = 1.10-2.25 for a dozen of bursts. These results are given 
on Table 1. See also Greiner (2000) for an updated information. About 50\% 
of the GRBs with X-ray counterparts are not detected in the optical, and this 
could be due to intrinsic faintness because of a low ambient medium, high 
absorption in a dusty enviroment, or Lyman limit absorption in high redshift 
galaxies (z $>$ 7).  Table 2 summarizes the properties of the host galaxies 
found so far.

\begin{table}[]
\hspace{1.5cm} 
\caption{GRBs detected by {\it BeppoSAX} and {\it RXTE} in 1996-99}
\begin{tabular}{llllllll}
\hline
GRB    & X-rays   &optical-IR& radio & GRB    & X-rays   &optical-IR& radio \\
\hline
960720 &          &          &       & 990123 &  yes     &   yes    & yes  \\
970111 &  yes ?   &   no     &  no   & 990217 &  no      &   no     &      \\
970228 &  yes     &   yes    &  no   & 990308 &  yes ?   &   yes    &      \\
970402 &  yes     &   no     &       & 990506 &  yes     &   no     & yes ?\\
970508 &  yes     &   yes    &  yes  & 990510 &  yes     &   yes    &      \\
970616 &  yes ?   &   no     &  no   & 990520 &  yes     &   no     & no   \\
970815 &  yes ?   &   no     &  no   & 990625 &          &          &      \\
970828 &  yes     &   no     &       & 990627 &  yes     &   no     &      \\
971214 &  yes     &   yes    &  no   & 990704 &  yes     &          &      \\
971227 &  yes ?   &   yes ?  &       & 990705 &  yes     &   yes    &      \\
980109 &          &   yes ?  &       & 990712 &          &   yes    &      \\
980326 &  yes ?   &   yes    &       & 990806 &  yes     &   no     &      \\
980329 &  yes     &   yes    &  yes  & 990907 &  yes ?   &   no     &      \\
980425 &  yes ?   &   yes ?  & yes ? & 990908 &          &   no     &      \\
980515 &  yes ?   &          &       & 991014 &  no      &   no     &      \\
980519 &  yes     &   yes    & yes   & 991105 &          &   no     &      \\
980613 &  yes     &   yes    &       & 991106 &  yes ?   &   no     &      \\ 
980703 &  yes     &   yes    &  yes  & 991208 &          &   yes    & yes  \\ 
980706 &  yes ?   &   no     &       & 991216 &  yes     &   yes    & yes  \\
981220 &  yes     &          &       & 991217 &          &   no     &      \\
981226 &  yes     &          & yes ? &        &          &          &      \\ 
\hline
\end{tabular} 
\end{table}

\begin{table}[]
\hspace{1.5cm} 
\caption{"classical" GRB host galaxies}
\begin{tabular}{llccl}
\hline
GRB    & R$_{host}$ &      $z$    &SFR ($M_{\odot}$ year$^{-1}$)&  References\\
\hline
970228 &  25.2    &         0.695        &  0.5  &  Djorgovski et al. (1999a)\\
980828 & 24.2 ?   &     0.33 ? 0.96 ?    &       &  Yoshida et al. (1999),   \\
       &          &                      &       &  Frail et al. (2000)      \\
970508 &  25.7    &         0.835        &   1   &  Bloom et al. (1998)      \\
971214 &  25.6    &         3.418        &   5   &  Kulkarni et al. (1998b)  \\
980329 &  26.3    &                      &       &  Djorgovski et al. (2000) \\
980519 & $\sim$26 &                      &       &  Hjorth et al. (1999)     \\
980613 &  23.8    &         1.096        &   3   &  Djorgovski et al. (1999b)\\
980703 &  22.5    &         0.966        &  63   &  Djorgovski et al. (1998) \\
981226 &  24.9    &                      &       &  Frail et al. (1999b)     \\
990123 &  23.8    &         1.599        &       &  Kulkarni et al. (99),    \\
       &          &                      &       &  Andersen et al. (1999)   \\
990506 &  24.8    &                      &       &  Frail et al. (2000)      \\
990510 & $\geq$27 &         1.619        &       &  Vreeswijk et al. (1999b) \\
990712 &  21.8    &         0.430        &       &  Galama et al. (1999b)    \\
991208 &   24     &         0.707        &       &  Dodonov et al. (1999)    \\
       &          &                      &       &  Djorgovski et al. (1999c)\\
991216 &   24     &         1.02         &       &  Vreeswijk et al. (1999c) \\
\hline
\end{tabular} 
\end{table}

\section{The relativistic blast wave model}

The observational characteristics of the GRB counterparts can be
accommodated in the framework of the relativistic fireball models, first 
proposed by Goodman (1986) and Paczy\'nski (1986), in which a compact source 
releases 10$^{53}$ ergs of energy within dozens of seconds in a region smaller
than 10 km. The opaque radiation-electron-positron plasma accelerates
to relativistic velocities (the fireball). The GRB itself are thought to be
be produced by a serie of "internal shocks" due to collisions amongst layers 
expelled with different Lorentz factors that are being caught up to each 
other. 
When the fireball runs into the surrounding medium, a "forward shock" 
ploughs into the medium, and sweeps up the interstellar matter, decelerating 
and producing an afterglow at frequencies gradually declining from X-rays 
to radio wavelenghts (M\'esz\'aros and Rees 1997). A "reverse 
shock" impinges on the ejecta. An extensive review is given by Piran (1999).

The properties of the blast wave can be derived from the classical synchrotron
spectrum (Ginzburg and Syrovatskii 1965) produced by a population of 
electrons with the addition of self absorption and 
a cooling break (Sari, Piran and Narayan 1998).
The determination for every GRB of the six observables:
the synchrotron, break and self-absorption frequencies, the maximum flux and 
the power-law decay exponent (all from the multiwavelength spectrum) 
and $z$ (from optical or X-ray spectroscopy) allows to obtain 
the total energy per solid angle, the fraction of the shock energy in 
electrons and post-grb magnetic fields, and the density of the ambient medium.

How does the GRB take place? 
The most popular model is that of a ``failed'' type-I SN (Bodenhaimer and
Woosley 1983, Woosley 1993) or {\it hypernova} (as it has been called by 
Paczy\'nski (1998) on the basis of the observational consequences): 
very massive stars (Wolf-Rayet) collapse forming a Kerr black hole (BH) 
and a 0.1-1 $M_{\odot}$ torus. 
The matter is accreted at a very high rate and the energy is
extracted via the rotational energy of the BH (Lee et al. 1999) 
or via the accretion energy from the disk. 
In any case, a ``dirty fireball'', is produced 
reaching a luminosity $\sim$ 300 times larger that than of a normal SN. 
This would happen every $\sim$ 10$^{6}$ yr. In this scenario, GRBs
would be produced in dense enviroments near star forming 
regions (see also MacFadyen and Woosley 1999) and GRBs might be used
for deriving the SFR in the Universe (Krumholz et al. 1998, Totani et al. 
1999).

The coalescence of neutron stars in a binary system has been also 
proposed (Narayan et al. 1992): lifes of such systems are of the order of
$\sim$ 10$^{9}$ years, and large escape velocities are usual, putting them 
far away from the regions where their progenitors were born. The likely 
result is a Kerr BH, and the energy released energy during 
the merger process is $\sim$ 10$^{54}$ erg. It is also possible that a
$\sim$ 0.1 $M_{\odot}$ accretion disk forms around the black hole and is 
accreted within a few dozen seconds, then producing internal shocks leading 
to the GRB (Katz 1997). There are variations of these models where one or
two components are substituted for black holes (Paczy\'nski 1991).
It fact, there it has been suggested that the short duration ($<$ 1 s) 
bursts could be due to compact star mergers, whereas the longer ones 
are caused by the collapse of massive stars.

\section{Summary}
The existence of X-ray afterglow in {\it most} bursts is confirmed.
Out ot 26 {\it BSAX} pointings, 17 revealed a clear afterglow, 
leading to the detection of several
optical/IR/radio counterparts in 1997-99. The determination of $z$ 
for the host galaxies by means of absorption edges or emission lines 
in the X-ray afterglow seems to be very promising. This requires prompt 
X-ray follow-up, as was achieved for GRB 970508 and GRB 970828.
However, only the population of bursts
with durations of few seconds has been explored. Short bursts lasting
less than 1 s, like GRB 980706, that follow the -3/2 slope in the 
log $N$-log $S$ diagram (in contrast to the longer bursts) remain to be 
detected at longer wavelengths.
Energy releases of $\sim$ 10$^{54}$ erg (as derived for GRB 980329 and 
990123) are difficult 
to reconcile with theoretical models and non-isotropic emission, such as 
intrinsic beaming appears as the most plaussible resolution of this problem.



\end{document}